\documentclass[twocolumn, amsmath, superscriptaddress, pra]{revtex4}
\usepackage{amssymb}
\usepackage{amsmath}
\usepackage{graphicx}
\usepackage{dcolumn}
\usepackage{bm}
\usepackage{tabularx}

\begin{document}

\title{Finite-size scaling analysis in the two-photon Dicke model}
\author{Xiang-You Chen}
\affiliation{Department of Physics, Chongqing University, Chongqing
401330, People's Republic of China}
\author{Yu-Yu Zhang}
\email{yuyuzh@cqu.edu.cn}
\affiliation{Department of Physics, Chongqing University, Chongqing
401330, People's Republic of China}

\date{\today }

\begin{abstract}
We perform a Schrieffer-Wolff transformation to the two-photon Dicke model
by keeping the leading-order correction with a quartic term of the field,
which is crucial for finite-size scaling analysis.
Besides a spectral collapse as a
consequence of two-photon interaction, the super-radiant phase
transition is indicated by the vanishing of the excitation energy and
the uniform atomic polarization.
The scaling functions for the
ground-state energy and the atomic pseudospin are derived analytically.
The scaling exponents of the observables
are the same as those in the standard Dicke model, indicating they are in the same universality class.
\end{abstract}

\maketitle

\section{Introduction}
The Dicke model~\cite{dicke} describes a collection of $N$ two-level atoms
interacting with a single radiation mode via an atom-field coupling. Due to
the spontaneous coherent radiation of the atomic ensemble, a super-radiant
quantum phase transition (QPT) occurs~\cite{emary,chen} in the ultra-strong coupling (USC) regime,
where the atom-field coupling strength is comparable to the field
frequency~\cite{ciuti,fumiki,Yoshihara,solano}.
There is ongoing interest in the realization of the super-radiant phase in
circuit quantum electrodynamics (QED) systems~\cite{jaako,bamba,Ashhab,Kakuyanagi},
where two-level qubits are strongly coupled to microwave cavities.
Such experimental achievement has prompted a number of theoretical
efforts for generalizations of the Dicke model, such as including anisotropic couplings~\cite{baksic,sprik,ye}
and two-photon interaction.~\cite{travenec,felicetti,garbe}.

In particular, two-photon interaction usually describes a second-order process
in different physical setups, such as Rydberg atoms in microwave
superconducting cavities~\cite{bertet,xue} and quantum dots~\cite%
{stufler,valle}. For an atom coupling to the field via two-photon interaction,
the interesting finding is a spectral
collapse, for which all discrete system spectrum collapse into a continuous band~\cite{wang,duan1,peng,lv}.
In a collective of atoms system described by the two-photon Dicke model, the important finding besides a
spectral collapse is a super-radiant phase transition~\cite{garbe},
which is induced by coherent radiations of the atoms.
However, the universal scaling and critical exponents of the super-radiant QPT in the two-photon Dicke model
remain elusive.
The finite-size correction in many-body system has been shown to be crucial in the understanding
of the universality class in the QPT~\cite{chen1,liu,vidal3,liberti,feng}. Numerically, it is very challenging
to give a convincing exact treatment
of the finite-size two-photon Dicke model. So it is highly desirable to explore
finite-size scaling exponents in the atomic ensemble, which are
significant for distinguishing the universality class.

The main motivation of this paper is to investigate the universal critical exponents
by the analytical scaling functions.
We employ a Holstein-Primakoff expansion~\cite%
{emary} and Schrieffer-Wolff (SW) transformation~\cite%
{plenio1,plenio2,loss,sanz} to diagonalize the Hamiltonian beyond the mean-field approximation.
In contrast to the mean-field analysis by including second-order quantum fluctuations~\cite{garbe},
a lower excitation energy is obtained in the super-radiant phase by our method.
Moreover, as an improvement, a quartic potential for the field is added to the leading-order
corrections to the effective Hamiltonian, which is crucial to study the quantum criticality.
Critical exponents of the ground-state energy and the atomic pesudospin
are extracted analytically from the universal finite-size scaling functions.
We show that the super-radiant QPT in the two-photon Dicke model belongs to the same
universality class as the standard Dicke model~\cite{vidal3,chen1}.

The paper is outlined as follows. In Sec.~II, the Hamiltonian is diagonalized by a Holstein-Primakoff
expansion and SW transformations in the normal and super-radiant phases, respectively. In Sec.~III, analytical
expressions for some observables are evaluated to show the super-radiant phase transition.
In Sec.~IV, we discuss the universal finite-size scaling in the critical regime, and the critical exponents are
given analytically. Finally, a brief summary is given in Sec.~V.

\section{Thermodynamic limit}

The Hamiltonian of the two-photon Dicke model, where $N$ identical two-level atoms interacting with a single bosonic
mode via two-photon interaction, is
\begin{equation}
H=\Delta J_{z}+\omega a^{\dagger }a+\frac{2g}{N}\left( a^{\dagger
2}+a^{2}\right) J_{x},  \label{Hamiltonian}
\end{equation}
where $a^{\dagger }$ $\left( a\right) $ is the creation
(annihilation) operator of the single-mode cavity with frequency $\omega $.
The colletive angular momentum
operators $J_{z}=\sum_{i=1}^{N}\sigma _{z}^{(i)}/2$ and $J_{x}=%
\sum_{i=1}^{N}\sigma _{x}^{(i)}/2$ describe the ensemble of $N$ two-level
atoms with a pseudospin $j=N/2$. And $\Delta $ is the atomic transition frequency,
and $g$ is the collective coupling strength of two-photon interaction.

The Hamiltonian commutes with a generalized $Z_{4}$ parity operator $\Pi $,
which is defined by $\Pi =(-1)^{N}\otimes _{n=1}^{N}\sigma
_{z}^{(n)}e^{i\pi a^{\dagger }a/2}$. $\Pi $ has four eigenvalues $\pm 1$
and $\pm i$, and is different from the $Z_{2}$ parity in the standard
Dicke model~\cite{emary,chen}. The $Z_{4}$ parity symmetry in the ground state
is expected to be spontaneously broken in the super-radiant phase transition.

It is convenient to describe two-photon interaction by introducing new operators $
K_{0}=\frac{1}{2}(a^{\dagger }a+\frac{1}{2})$, $K_{+} =\frac{1}{2}a^{\dagger 2}$,
and $K_{-}=\frac{1}{2}a^{2}$, which form the $SU(1,1)$ Lie algebra and obey commutation
relations $[K_{0},K_{\pm }]=\pm K_{\pm }$, and $[K_{+},K_{-}]=-2K_{0}$. Then, we use the
Holstein-Primakoff transformation of the collective angular momentum operators defined as $
J_{+}=b^{\dagger }\sqrt{N-b^{\dagger }b}$, $J_{-}=\sqrt{N-b^{\dagger }b}b$,
and $J_{z}=b^{\dagger }b-N/2$ with $[b,b^{\dagger }]=1$. After that, the Hamiltonian takes the form
\begin{eqnarray}  \label{ham1}
H &=&\Delta (b^{\dagger }b-N/2)+\omega (2K_{0}-\frac{1}{2})  \notag \\
&&+\frac{2g}{\sqrt{N}}\left( K_{+}+K_{-}\right) (b^{\dagger }\sqrt{1-\frac{%
b^{\dagger }b}{N}}+\sqrt{1-\frac{b^{\dagger }b}{N}}b).\notag \\
\end{eqnarray}
We consider the two-photon Dicke model in the
thermodynamic limit for infinite atoms $N\rightarrow \infty $.
By means of the boson expansion approach,we expand the Hamiltonian with respected to
the bosonic operator $b^{\dagger}$($b$) as power series in $1/N$.

\subsection{Normal phase}
We derive the Hamiltonian of the normal phase by simply neglecting terms of
order O($1/N^{3/2}$) in Eq.(~\ref{ham1}) as
\begin{eqnarray}
H_{np}=\frac{\omega_1}{N} b^{\dagger }b+2\omega K_{0}+\lambda(b^{\dagger }+b)\left( K_{+}+K_{-}\right)-\frac{\omega+\omega_1}{2},\notag\\
\end{eqnarray}%
where the parameters $\omega_1=N\Delta$ and $\lambda =2g/\sqrt{N}$.

Inspired by the SW transformation~\cite{plenio1,plenio2,loss,sanz}, we present a treatment of $H_{np}$ basing on the unitary transformation $U=e^{R}$ with the generator $R=\lambda R_{1}+\lambda ^{3}R_{3}$. The aim of the SW transformation
is to eliminate the block-off-diagonal interacting terms,
such as $(b^{\dagger }+b)\left( K_{+}+K_{-}\right)$, and to keep the block-diagonal coupling terms such as $(b+b^{\dagger
})^{2}K_{0}$(see Appendix A).
Consequently, we keep the terms up to order $1/N^{2}$ and the higher order terms can be
neglected. It results in the transformed Hamiltonian $H_{np}^{\prime }=H_{1}^{\prime
}+H_{2}^{\prime }$, consisting of
\begin{equation}\label{H1}
H_{1}^{\prime }=\frac{\omega_1}{N} b^{\dagger }b-\frac{4g^{2}}{N\omega }(b+b^{\dagger
})^{2}K_{0}+2\omega K_{0}-\frac{\omega +\omega_1}{2},
\end{equation}
and
\begin{equation}\label{H2}
H_{2}^{\prime }=-\frac{4g^{4}}{N^{2}\omega ^{3}}(b+b^{\dagger })^{4}K_{0}-%
\frac{\omega_1g^{2}}{N^2\omega ^{2}}(K_{+}-K_{-})^{2}.
\end{equation}
The Hamiltonian is free of coupling terms between $b+b^{\dagger }$ and $K_{+}-K_{-}$,
and can be simply diagonalized in the subspace of $K_{0}$ with $\langle K_0\rangle=1/4$.
Especially, the terms $H_{2}^{\prime }$ involves a quartic potential for the field,
which plays a crucial role in the finite-size scaling ansatz. Eq.(~\ref{H1}) can be diagonalized
to be $H_{np}=\varepsilon _{1}(g)b^{\dagger }b+E_{g}^{(1)}$
by a squeezing operator $S=e^{\zeta (b^{2}-b^{+2})/2}$ with $\zeta =-\mathtt{ln}(1-\frac{4g^{2}}{N\omega \Delta })/4$.
And the excitation energy is obtained as $\varepsilon _{1}(g)=\omega_1\sqrt{1-g^{2}/g_{c}^{2}}/N$,
which is real only when $g\leqslant \sqrt{\omega \omega_1}/2=g_{c}$.
With the inclusion of the term $H_{2}^{\prime }$, the ground-state energy in the normal phase is
\begin{eqnarray}
E_{g}^{(1)} &=&-\frac{\omega_1}{2}+\frac{\omega_1 }{2N}(\sqrt{1-\frac{g^{2}}{%
g_{c}^{2}}}-1) \notag \\
&&-\frac{g^{2}}{N^2}[\frac{\omega _{1}}{2\omega ^{2}} +\frac{g^{2}g_{c}^{2}}{\omega ^{3}(g_{c}^{2}-g^{2})}].  \label{egnp}
\end{eqnarray}
By comparing with the mean-field results~\cite{garbe}, the ground-state energy is obtained by keeping terms of order $1/N^2$.
Meanwhile, the ground state for the normal phase is
$|\varphi_{np}\rangle=U^{\dag}S^{\dag}|0\rangle_{b}|0\rangle_{K_0}$, where $|0\rangle_b$
is the vacuum state of the atom ensemble and $|0\rangle_{K_0}$ is the ground state of $K_0$.
One can easily obtain the expectation value of the bosonic operator $\langle\hat{b}\rangle$, which equals to zero in the normal phase.

\subsection{Superradiant Phase}
In the super-radiant phase, there occurs a uniform atomic polarization and the pseudospin $J_z$ is
polarized along the $z$ axis. In the Holstein-Primakoff representation, the atomic operator $b$ is
expected to be shifted as
\begin{equation}  \label{beta}
d=D^{\dagger}[-\beta\sqrt{N}]bD[-\beta\sqrt{N}]=\beta \sqrt{N}+b,
\end{equation}
with a unitary transformation $D[-\beta\sqrt{N}]=e^{-\beta\sqrt{N}(\hat{b}^{\dagger}-\hat{b})}$.
As previously reported, the displacement $\beta$ is obtained by the mean-field value~\cite%
{garbe}. We proceed to determine the variable $\beta$ beyond the mean-field
approximation.

Due to the shifted displacement of $b$, it is obvious that the expectation value of $b$ in the super-radiant state
is $\beta \sqrt{N}$.
Whereas the displacement of the field operator $a$ equals to zero due to the
absence of linear interactions between atoms and cavity. As a consequence,
the Hamiltonian of Eq.(~\ref{ham1}) becomes%
\begin{eqnarray}\label{hsp}
H_{sp} &=&\frac{\omega}{N}d^{\dagger }d+\frac{\omega_1}{\sqrt{N}} \beta (d^{\dagger }+d)+\frac{%
2g\beta _{1}\beta _{2}}{\sqrt{N}}(d^{\dagger }+d)\left( K_{+}+K_{-}\right)
\notag \\
&&-\frac{g\beta }{N\beta _{1}}[d^{\dagger 2}+d^{2}+4d^{\dagger }d]\left(
K_{+}+K_{-}\right) \notag \\
&&+H_f+\beta _{0}+O(N^{-3/2}),
\end{eqnarray}%
where the field part in the Hamiltonian is $H_f=2\omega K_{0}+\lambda_{\beta }\left( K_{+}+K_{-}\right)$,
and the parameters are given by $\beta _{1}=\sqrt{1-\beta ^{2}}$, $\beta _{2}=1-\beta ^{2}/(1-\beta
^{2})$ , $\beta _{0}=\omega_1\beta ^{2}-(\omega_1+\omega )/2$ and $\lambda
_{\beta }=4g\beta \beta _{1}$.

Firstly, we apply a squeezing operator $S[r]=e^{-r(a^{\dagger 2}-a^{2})/2}$ to diagonalize the field part of the above Hamiltonian $H_f$.
And the transformed Hamiltonian is derived as $H_{2}^{(0)}+V_{1}+V_{2}+V_{3}+V_{linear}$ in Appendix B.
We now choose the displacement $\beta$ to eliminate the term $V_{linear}$ in Eq.(~\ref{vlinear}) that is linear in the bosonic operators.
It gives
\begin{equation}
\omega_1 \beta -g\beta _{1}\beta _{2}\frac{\lambda _{\beta }}{\sqrt{\omega
^{2}-\lambda _{\beta }^{2}}}=0.
\end{equation}%
The $\beta=0$ solution recovers the normal phase Hamiltonian. The nontrivial solution gives
\begin{equation}
\beta =\frac{1}{\sqrt{2}}\left[1-\sqrt{\frac{1-4g^{2}/\omega ^{2}}{%
16g^{4}/(\omega\omega_1)^{2}-4g^{2}/\omega ^{2}}}\right]^{1/2},
\end{equation}%
which remains real, provided that $1-4g^{2}/\omega ^{2}\geqslant 0$ and $1-%
\sqrt{\frac{1-4g^{2}/\omega ^{2}}{16g^{4}/(\omega\omega_1
)^{2}-4g^{2}/\omega ^{2}}}\geqslant 0$. It leads to the collapse point and the
critical value
of coupling strength, respectively,
\begin{equation}
g_{collapse}=\omega /2,
\end{equation}%
and
\begin{equation}
g_{c}=\frac{\sqrt{\omega \omega_1}}{2}.
\end{equation}

Our solutions shows that the super-radiant QPT occurs at the critical point $g_{c}$,
which is characterized by nonvanishing of the expectation value of $b$.
Interestingly, the spectrum collapses at $g_{collapse}$,
so that the Hamiltonian is not bounded from below and the model is not
well defined. We focus on the parameter regime where the phase transition
can be accessed in the validity coupling region $g<g_{collapse}$.
Moreover, since the super-radiant phase transition occurs before the
spectral collapse, one have the condition $\omega_1=N\Delta <\omega $, requiring that
the order of magnitude of $\Delta $ is $\omega /N$. Hence the scaled atom
frequency $\omega _{1}=N\Delta $ is introduced and is comparable to the field frequency $%
\omega $.

Then, by eliminating the block-off-diagonal coupling terms $V_1$ in Eq.(~\ref{v1}) and $V_2$ in Eq.(~\ref{v2}),
the Hamiltonian in the super-radiant phase
$H_{sp}$ can be diagonalized as
\begin{equation}
H_{sp}=\varepsilon _{2}(g)(d^{\dagger }d+\frac{1}{2})+E_{g}^{(2)},
\end{equation}%
where the excitation energy is%
\begin{equation}  \label{esp}
\varepsilon _{2}(g)=\frac{2\omega _{1}-\lambda _{3}}{2N}\sqrt{1-\frac{%
2\lambda _{1}^{2}/(2\sqrt{\omega ^{2}-\lambda _{\beta }^{2}}+\lambda
_{3}/N)+\lambda _{3}}{(\omega _{1}-\lambda _{3}/2)}},
\end{equation}%
and the ground-state energy is
\begin{equation}  \label{egsp}
E_{g}^{(2)}=\frac{1}{2}\varepsilon _{2}(g)-\frac{\omega _{1}-\lambda _{3}}{2N%
}+\frac{\sqrt{\omega ^{2}-\lambda _{\beta }^{2}}}{2}+\beta _{0}
\end{equation}
with the parameters $\lambda _{1}$ and $\lambda _{3}$ in the Appendix B.
Thus, we obtain the diagonal Hamiltonian $H_{sp}$ for the
super-radiant phase. If we choose the signs of the displacement as $-\beta$ in Eq.(~\ref{beta}), we obtain an identical
effective Hamiltonian. It is clear that the spectrum is doubly degenerate
in the super-radiant phase.

\section{Phase transition}

\begin{figure}[tbp]
\includegraphics[scale=0.75]{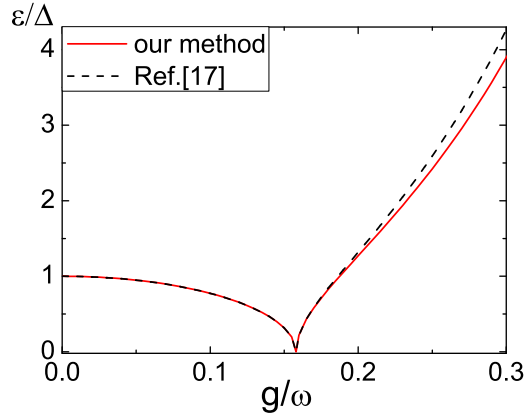}
\caption{Excitation energy $\protect\varepsilon (g)/\Delta $ obtained by our
method (red solid line) as a function of coupling $g/\protect\omega $ for $%
\omega_1=0.1\omega$. For comparison, results obtained by mean-field
analysis in Ref.~\protect\cite{garbe} (black dashed line) are
calculated.}
\label{excitation}
\end{figure}

After deriving the two effective Hamiltonian in the $N\rightarrow \infty $
limit, we now explore the properties of two phases. The excitation energies are
given by $\varepsilon _{1}(g)$ in the normal phase and $\varepsilon _{2}(g)$
in the super-radiant phase. Fig.~\ref{excitation} displays the behavior of
the excitation energies as a function of coupling strength $g/\omega $,
which is lower than the mean-field result~\cite{garbe} in the super-radiant phase.
As the coupling approaches the critical
value $g\rightarrow g_{c}$, the excitation energy can be shown to vanish as
\begin{equation}
\varepsilon (\lambda \rightarrow \lambda _{c})\sim \frac{\omega _{1}}{N}%
\sqrt{\frac{2}{g_{c}}}(g_{c}-g)^{1/2}.
\end{equation}
The vanishing of the excitation energies at $g_c$ reveals that
second-order phase transition occurs.

\begin{figure}[tbp]
\includegraphics[scale=0.75]{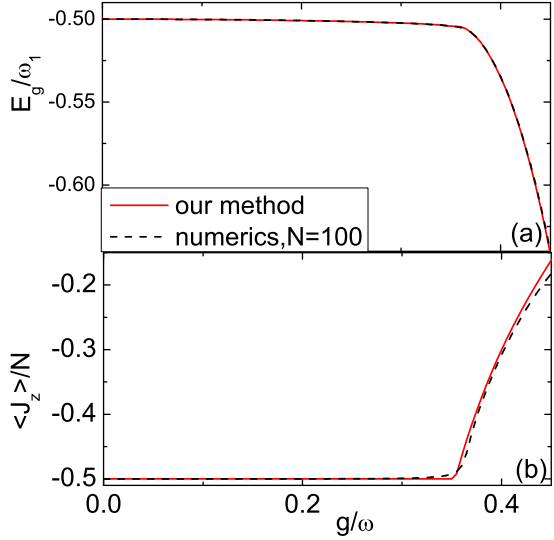}
\caption{The scaled ground-state energy $E_g/(\omega_1)$ (a) and
the expected value of the scaled pesudospin $J_z/N$(b)
obtained by our method as a function of coupling $g/\protect\omega $ for $
N=100 $ and $\omega_1/\omega=0.5$. Solid lines denote our analytical results,
whereas dashed lines correspond to exact-diagonalization ones.}
\label{egjz}
\end{figure}

Fig.~\ref{egjz}(a) shows the scaled ground-state energy for the normal and super-radiant phases
according to the analytical expression in Eqs.(~\ref{egnp}) and (~\ref{egsp}), which
are in consistent with the numerical ones for $N=100$ atoms.
In the thermodynamic limit $N\rightarrow\infty$, the scaled ground-state energy $E_g/\omega_1$ at the
critical point $g_c$ equals to $-1/2$, as shown in Table~\ref{table}.

We calculate the expectation value of the scaled pseudospin
\begin{equation}
\langle J_{z}\rangle/N =\beta ^{2}-1/2.
\end{equation}
It makes clear the physical meaning of the displacement parameter $\beta $
in Eq.(~\ref{beta}), which illustrates the uniform atomic polarization along $z$ axis.
In Fig.~\ref{egjz}(b), $\langle J_{z}\rangle/N $ becomes larger than $-1/2$ when
the coupling strength exceeds the critical point $g_c=\sqrt{2}\omega/4$ for $\omega_1/\omega=0.5$.

As demonstrated above, the behavior of the excitation energies $\varepsilon(g)$,
the scaled ground-state energy $E_g/\omega_1$ and the pseudospin
$\langle J_{z}\rangle /N$ are similar to those in the standard Dicke model
in the thermodynamic limit~\cite{emary,chen}. It becomes interesting
to explore the critical exponents and universality class of the two-photon Dicke model.

\section{Finite-size scaling}
It is well know that different systems can exhibit similar behavior in the critical regime, giving rise to the universality.
Finite-size scaling is a topic of major interest in the QPT
system and has solid foundations since the formulation of a general theory~%
\cite{fisher,botet}. As shown in previous studies~\cite{vidal1,vidal2,vidal3}%
, the $1/N$ corrections to physical observables such as order parameters
display some singularities at the critical point. We now proceed to
derive finite-size scaling functions analytically for some observables in the two-photon Dicke model.

We start with the Hamiltonian $H_{np}^{\prime }=H_{1}^{\prime }+H_{2}^{\prime }$ in Eqs.(~\ref{H1}) and (~\ref{H2}),
by including the quartic term for the field. By projecting the Hamiltonian to
the subspace $|0\rangle_{K_{0}}$, we obtain
\begin{eqnarray}
H_{np}^{\prime } &=&\frac{\omega _{1}}{N}b^{\dagger }b-\frac{\omega _{1}g^{\prime 2}}{%
4N}(b+b^{\dagger })^{2}-\frac{\omega _{1}^{2}g^{\prime 4}}{16N^{2}\omega }%
(b+b^{\dagger })^{4}+c,  \notag \\
\end{eqnarray}
where $g^{\prime }=g/g_{c}$ and a constant term $c=-\omega _{1}/2-\omega _{1}/(2N)-\omega _{1}g^{2}/
(2N^{2}\omega ^{2})$. To understand the properties of
the phase transition, we rewrite $H_{np}^{\prime }$ by the introduction of coordinate
and momentum operators for the bosonic mode, $x=1/\sqrt{2\omega _{1}/N}%
(b^{\dagger }+b)$ and $p=i\sqrt{\frac{\omega _{1}}{2N}}(b^{\dagger }-b)$, as follows
\begin{equation}
H_{np}^{\prime }=\frac{1}{2}p^{2}+\frac{\omega _{1}^{2}}{2N^{2}}(1-g^{\prime 2})x^{2}-%
\frac{\omega _{1}^{4}g^{\prime 4}}{4N^{4}\omega }x^{4}-\frac{\omega _{1}}{2N}.
\end{equation}
It is helpful to rescale the coordinate by $x=\tilde{x}N^{\alpha }$ and the
corresponding momentum $p=-i\partial /\partial x=\tilde{p}N^{-\alpha }$.
Then the Hamiltonian becomes%
\begin{equation}
H_{np}^{\prime }=\frac{1}{2}\tilde{p}^{2}N^{-2\alpha }+\frac{\omega _{1}^{2}}{2}%
(1-g^{\prime 2})\tilde{x}^{2}N^{2\alpha -2}-\frac{\omega _{1}^{4}g^{\prime 4}%
}{4\omega }\tilde{x}^{4}N^{4\alpha -4}.
\end{equation}
By setting $\alpha =2/3$, we obtain the scaling variable
\begin{equation}\label{scaling}
\eta =\frac{\omega _{1}^{2}}{2}(1-g^{\prime
2})N^{2/3}
\end{equation}
and $\tilde{x}=xN^{-2/3}$. The renormalized Hamiltonian is written as
\begin{equation}
H_{np}^{\prime }=N^{-4/3}[-\frac{\partial ^{2}}{2\partial \tilde{x}^{2}}+\eta \tilde{x}%
^{2}-\frac{\omega _{1}^{4}g^{\prime 4}}{4\omega }\tilde{x}^{4}],
\end{equation}
which is crucial to reveal the universal properties of the second-order QPT.

\begin{table}[tbp]
\caption{finite-size scaling exponents for the ground-state energy $E_g/%
\protect\omega_1$, the scaled atomic angular momentum $\langle J_z\rangle/N$
and $\langle J_y^2\rangle/N^2 $ for the two-photon Dicke model. We find that
the corresponding scaling exponents are the same as those in the standard Dicke
model~\cite{vidal3,chen1}.}
\label{table}\centering\label{table}
\begin{tabular}{ccc}
\hline
$Q_N$ & $\lim_{N\rightarrow\infty}Q_N$ & two-photon Dicke
\\ \hline \\
$E_g/\omega_1$ & -1/2 & -4/3 \\ \\
$\langle J_z\rangle/N$ & -1/2  & -2/3  \\\\
$\langle J_y^2\rangle/N^2$ & 0 & -4/3 \\ \hline
\end{tabular}%
\end{table}

\begin{figure}[tbp]
\includegraphics[scale=0.65]{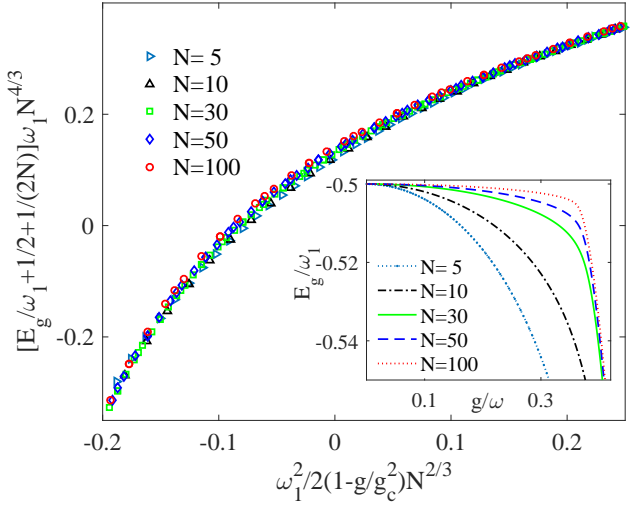}
\caption{Finite-size scaling for the scaled ground-state energy
in the two-photon Dicke model.
Points corresponding to different $N$ collapse on the same curve.
Inset: the ground-state energy $E_g/\omega _{1}$ as a function of the coupling strength $g/\omega$ for different $N$.}
\label{scaledenergy}
\end{figure}

The ground-state wavefunction $\varphi _{0}(\tilde{x},\eta )$ is described straightforwardly
by the following equation in terms of $\tilde{x}$ and $\eta $:%
\begin{equation}
\lbrack -\frac{\partial ^{2}}{2\partial \tilde{x}^{2}}+\eta \tilde{x}^{2}-%
\frac{\omega _{1}^{4}g^{\prime 4}}{4\omega }\tilde{x}^{4}]\varphi _{0}(%
\tilde{x},\eta )=E_{0}(\eta )\varphi _{0}(\tilde{x},\eta ),
\end{equation}
where $E_{0}(\eta )$ gives the ground-state energy as
\begin{equation}  \label{eg}
E_{g}=-\frac{\omega _{1}}{2}-\frac{\omega _{1}}{2N}+\frac{1}{N^{4/3}}E_{0}(\eta).
\end{equation}
From the leading-order correction for the ground-state energy in the above equation,
the finite-size scaling exponent of $E_{g}$ is found to be $-4/3$, which is
the same as that for the Dicke model~\cite{chen1,vidal3}, as shown in Table~\ref{table}.

Meanwhile, the scaling law of the atomic ensemble
angular momentum $\langle J_{z}\rangle /N$ $=\langle b^{\dagger
}b-N/2\rangle /N$ and $\langle J_{y}^{2}\rangle /N^{2}$ can be derived as%
\begin{eqnarray}  \label{jz}
\langle J_{z}\rangle /N &=&-\frac{1}{2}+\frac{\omega _{1}}{2}N^{-2/3}X(\eta)
+\frac{1}{2\omega _{1}}N^{-4/3}P(\eta),\notag \\
\end{eqnarray}
and
\begin{equation}  \label{jy2}
\langle J_{y}^{2}\rangle /N^{2}=\frac{1}{2\omega _{1}}N^{-4/3}P(\eta),
\end{equation}
where the universal functions $X(\eta)$
and $P(\eta)$ are the expectation values of $\tilde{x}^2$ and $\tilde{p}^2$
over the ground state $\varphi(\tilde{x},\eta)$.
One can see that the leading-order finite-size corrections for  $\langle J_{z}\rangle /N$ and $\langle
J_{y}^{2}\rangle /N^{2}$ scale as  $N^{-2/3}$ and $N^{-4/3}$, respectively.
The finite-size scaling exponents are identical to those in the
standard Dicke model~\cite{chen1,vidal3} in Table~\ref{table}, providing an
evidence of the same universality class.

In general, the $1/N$ expansion of a physical quantity $Q_{N}(g)$ in the
vicinity of the critical point of the QPT, can be decomposed in a regular
and a singular function as follows~\cite{vidal2}:%
\begin{equation}
Q_{N}(g)=Q_{N}^{\mathtt{reg}}(g)+Q_{N}^{\mathtt{sing}}(g),
\end{equation}%
where $Q_{N}^{\mathtt{reg}}(g)$ and $Q_{N}^{\mathtt{sing}}(g)$ are regular and
singular functions at $g=g_{c}$. With the scaling variable $\eta$ in Eq.(~\ref{scaling}),
the singular function for an observable in the two-photon Dicke model is given explicitly as
\begin{equation}
Q_{N}^{\mathtt{sing}}(g)=F_{Q}[\frac{\omega _{1}^{2}(1-g^{2}/g_{c}^2)}{2}%
N^{2/3}],
\end{equation}%
where $F_{Q}$ is a scaling function depending only on the scaling variable $\omega _{1}^{2}(1-g^{2}/g_{c}^2)N^{2/3}/2$.

\begin{figure}[tbp]
\includegraphics[scale=0.65]{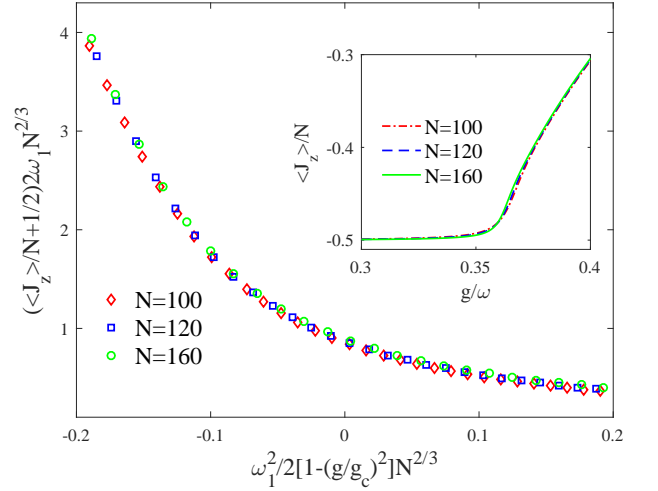}
\caption{Finite-size scaling for the scaled pesudospin $\langle J_z\rangle/N$ in the two-photon Dicke model.
Points corresponding to different $N$ collapse on the same curve.
Inset: $\langle J_z\rangle/N$ as a function of the coupling strength $g/\omega$ for different $N$.}
\label{scaledJz}
\end{figure}

\begin{figure}[tbp]
\includegraphics[scale=0.65]{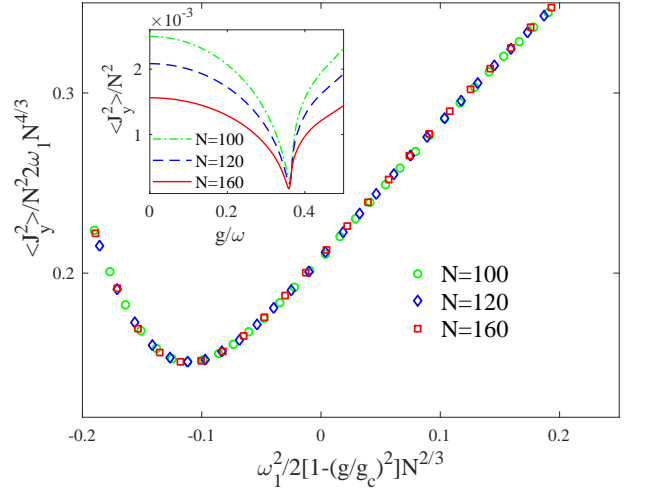}
\caption{Finite-size scaling for the scaled pesudospin $\langle J_y^2\rangle/N^2$ in the two-photon Dicke model.
Points corresponding to different $N$ collapse on the same curve.
Inset: $\langle J_y^2\rangle/N^2$ as a function of the coupling strength $g/\omega$ for different $N$.}
\label{scaledJy}
\end{figure}

Fig.~\ref{scaledenergy} shows the finite-size scaling for the scaled ground-state energy
for different sizes $N=5$, $10$, $30$, $50$ and $100$. The singular part of the ground-state energy $E_g+\omega_1/(2N)+\omega_1/(2N^2)$
for different sizes all collapse into a single curve in the critical regime. The numerical results
confirm the validity of the universal function $E_{0}(\eta )$ in Eq.(\ref{eg}), which is independent on $N$.
We also calculate the singular part of
$\langle J_{z}\rangle /N+1/2$ in Fig.~\ref{scaledJz} and $\langle J_y^2\rangle/N^2$
in Fig.~\ref{scaledJy}.
Excellent collapses in the critical regime are also achieved.
The numerical scaling results agree with the universal scaling functions $X(\eta)$ in Eq.(\ref{jz})
and $P(\eta)$ in Eq.(\ref{jy2}). The above results demonstrate that the finite-size
scaling functions by our treatment capture the universal laws of different observables.

\section{Conclusions}

In this paper, by combining the Schrieffer-Wolff transformation with
the Holstein-Primakoff expansion,
we diagonalize the Hamiltonian of the two-photon Dicke
model in the normal and super-radiant phases in thermodynamic limit, respectively.
In the super-radiant phase, the uniform atomic polarization is characterized
by the nonzero displacement of the atomic operator,
which is obtained beyond the mean-field approximation.
The vanishing of the excitation energy
at the critical coupling strength
illustrates the second-order super-radiant phase transition.

Since a convincing exact treatment of the finite-size two-photon Dicke model is lacking.
Our approach provides an efficient technique to derive the Hamiltonian by keeping the
leading-order correction with the quartic term for the field.
Consequently, the leading-order corrections and universal scaling functions for the
ground-state energy and the atomic angular momentum are derived analytically,
giving the finite-size scaling exponents precisely. We find that the two-photon Dicke model
and standard Dicke model are in the same universality class of QPT.

\acknowledgments
We acknowledge Qing-Hu Chen, Mao-Xin Liu and Li-Wei Duan for helpful discussion.
This work was supported by the Chongqing Research Program of Basic Research
and Frontier Technology (Grant No.cstc2015jcyjA00043 and
No.cstc2017jcyjAX0084 ).

\appendix

\section{Derivation of the effective Hamiltonian in the normal phase}
The Hamiltonian in the normal phase is written as $H_{np}=H_{0}+\lambda V$, consisting of
\begin{eqnarray}
H_{0} &=&\Delta b^{\dagger }b+2\omega K_{0}-\frac{\omega +\Delta N}{2}, \\
V &=&(b^{\dagger }+b)\left( K_{+}+K_{-}\right).
\end{eqnarray}

We consider a unitary transformation $U=e^{R}$ with the generator $R=\lambda R_{1}+\lambda
^{3}R_{3}$. The transformed Hamiltonian $H_{np}^{\prime }=e^{-R}H_{np}e^{R}$ is written as
\begin{eqnarray}
H_{np}^{\prime } &=&H_{0}+\lambda V+\lambda \lbrack H_{0},R_{1}]+\frac{%
\lambda ^{2}}{2}[V,R_{1}]  \notag \\
&&+\lambda ^{3}\{[H_{0},R_{3}]+\frac{1}{3}[[V,R_{1}],R_{1}]\}  \notag \\
&&+\lambda ^{4}\{\frac{1}{2}[V,R_{3}]-\frac{1}{24}[[[V,R_{1}],R_{1}],R_{1}]%
\}.
\end{eqnarray}
According to the SW transformation, the off-diagonal
coupling terms such as $V$ are required to be eliminated. One obtain
\begin{eqnarray}
\lbrack H_{0},R_{1}] &=&-V, \\
\lbrack H_{0},R_{3}] &=&-\frac{1}{3}[[V,R_{1}],R_{1}],
\end{eqnarray}
And the generators are determined as
\begin{eqnarray}
R_{1} &=&-\frac{1}{2\omega }(b+b^{\dagger })(K_{+}-K_{-}), \\
R_{3} &=&-\frac{1}{6\omega ^{3}}(b+b^{\dagger })^{3}(K_{+}-K_{-}).
\end{eqnarray}
Making use of the choice for the generators $R_1$ and $R_3$, the transformed Hamiltonian
becomes%
\begin{eqnarray}
H_{np}^{\prime } &=&\Delta b^{\dagger }b-\frac{4g^{2}}{N\omega }%
(b+b^{\dagger })^{2}K_{0}  \notag \\
&&-\frac{\Delta g^{2}}{N\omega ^{2}}(K_{+}-K_{-})^{2}-\frac{4g^{4}}{%
N^{2}\omega ^{3}}(b+b^{\dagger })^{4}K_{0}  \notag \\
&&+2\omega K_{0}-\frac{\omega +N\Delta}{2}+O(\frac{1}{N\sqrt{N}}).
\end{eqnarray}

\section{Derivation of the effective Hamiltonian in the super-radiant phase}

Let us now consider the Hamiltonian $H_{sp}$ in Eq.(~\ref{hsp}) in the super-radiant phase.
Firstly, the field part of the Hamiltonian $H_f=\lambda _{\beta }/2\left( a^{\dagger 2}+a^{2}\right) +\omega
(a^{\dagger }a+1/2)$ can be easily diagonalized by a squeezing transformation $S[r]=e^{-r(a^{\dagger
2}-a^{2})/2}$. It leads to
\begin{eqnarray}
S[r]H_fS^{\dagger}[r]&&=[\omega \cosh 2r+\lambda _{\beta }\sinh 2r](a^{\dagger }a+\frac{1}{2})
\notag \\
&&+\frac{1}{2}[\omega \sinh 2r+\lambda _{\beta }\cosh 2r]\left( a^{\dagger
2}+a^{2}\right).\notag \\
\end{eqnarray}
The squeezing parameter $r$ is determined by the vanishing of the $a^{\dagger 2}+a^{2}$ terms
\begin{equation}  \label{r}
r=\frac{1}{4}\texttt{ln}\frac{\omega -\lambda _{\beta }}{\omega +\lambda _{\beta }}.
\end{equation}
We perform the squeezing transformation to the Hamiltonian $H_{sp}$ in Eq.(~\ref{hsp}) as $S[r]H_{sp}S^{\dagger
}[r]=H_{2}^{(0)}+V_{linear}+V_{1}+V_{2}+V_{3}$. They are
\begin{eqnarray}  \label{H20}
H_{2}^{(0)} =\frac{\omega _{1}-\lambda _{3}K_{0}}{N}d^{\dagger }d+(2\sqrt{%
\omega ^{2}-\lambda _{\beta }^{2}}+\frac{\lambda _{3}}{N})K_{0}+\beta _{0},
\notag \\
\end{eqnarray}
\begin{equation}
V_{linear} =\frac{1}{\sqrt{N}}[\omega _{1}\beta +4g\beta _{1}\beta
_{2}\sinh (2r)K_{0}](d^{\dagger }+d),\\  \label{vlinear}
\end{equation}
\begin{equation}
V_{1} =\frac{\lambda _{1}}{\sqrt{N}}(d^{\dagger }+d)\left(
K_{+}+K_{-}\right) , \\ \label{v1}
\end{equation}
\begin{equation}
V_{2} =-\frac{\lambda _{2}}{N}(4d^{\dagger }d+d^{\dagger 2}+d^{2})\left(
K_{+}+K_{-}\right) , \\ \label{v2}
\end{equation}
\begin{equation}
V_{3} =-\frac{\lambda _{3}}{N}(d^{\dagger }+d)^{2}K_{0},
\end{equation}
where $\lambda _{1}=2g\beta _{1}\beta _{2}\cosh (2r)$, $\lambda _{2}=g\beta
\cosh (2r)/\beta _{1}$ and $\lambda _{3}=2g\beta \sinh (2r)/\beta _{1}$.
Here, we choose the value of $\beta$ to make the linear term $V_{linear}$ vanish.
Then, we employ a transformation $U=e^{\frac{1}{\sqrt{N}}P+\frac{1}{N}Q}$ with
the generators $P$ and $Q$ to
eliminate the block-off-diagonal terms $V_1$ and $V_2$. It leads to
\begin{eqnarray}
\frac{1}{\sqrt{N}}[H_{2}^{(0)},P] &=&-V_{1}, \\
\frac{1}{N}[H_{2}^{(0)},Q] &=&-V_{2},
\end{eqnarray}
which give the generators as
\begin{eqnarray}
P &=&-\frac{\lambda _{1}}{2\sqrt{\omega ^{2}-\lambda _{\beta }^{2}}+\lambda
_{3}/N}(d^{\dagger }+d)\left( K_{+}-K_{-}\right) , \\
Q &=&\frac{\lambda _{2}}{2\sqrt{\omega ^{2}-\lambda _{\beta }^{2}}+\lambda
_{3}/N}(4d^{\dagger }d+d^{\dagger 2}+d^{2})\left( K_{+}-K_{-}\right) .
\notag \\
\end{eqnarray}
After that, the transformed Hamiltonian becomes
\begin{eqnarray}
H_{sp}^{\prime }&=&\frac{1}{N}(\omega _{1}-2\lambda _{3}K_{0})d^{\dagger }d+(2\sqrt{%
\omega ^{2}-\lambda _{\beta }^{2}}+\frac{\lambda _{3}}{N})K_{0}  \notag \\
&&-\frac{1}{N}(\frac{2\lambda _{1}^{2}}{2\sqrt{\omega ^{2}-\lambda _{\beta
}^{2}}+\lambda _{3}/N}+\lambda _{3})(d^{\dagger }+d)^{2}K_{0}+\beta _{0}.\notag \\
\end{eqnarray}
By applying a squeezing transformation $S[r_{1}]=\exp [r_{1}^{2}(d^{\dagger
2}-d^{2})/2]$, we have
\begin{eqnarray}
H_{sp}^{\prime\prime }&=&S^{\dagger }[r_{1}]H_{sp}^{\prime }S[r_{1}]  \notag \\
&=&\frac{1}{N}[(\omega _{1}-2\lambda _{3}K_{0})\cosh 2r_{1}  \notag \\
&&-2(\frac{2\lambda_{1}^{2}}{2\sqrt{\omega ^{2}-\lambda _{\beta }^{2}}
+\lambda _{3}/N}+\lambda_{3}) e^{2r_{1}}K_{0}](d^{\dagger }d+\frac{1}{2})\notag \\
&&-\frac{\omega_{1}-2\lambda _{3}}{2N}
+(2\sqrt{\omega ^{2}-\lambda _{\beta }^{2}}+\frac{\lambda _{3}}{N})K_{0}+\beta _{0}\notag \\
&&+\lambda_4(d^{\dagger 2}+d^{2})
\end{eqnarray}
with $\lambda_4=\frac{1}{2N}[(\omega _{1}-2\lambda _{3}K_{0})\sinh 2r_{1}-2(\frac{%
2\lambda _{1}^{2}}{2\sqrt{\omega ^{2}-\lambda _{\beta }^{2}}+\lambda _{3}/N}
+\lambda _{3})e^{2r_{1}}K_{0}]$.
Making the $(d^{\dagger 2}+d^{2})$ term vanish in the subspace $|0\rangle_{K_{0}}$, we obtain the squeezing parameter
\begin{equation}  \label{r1}
r_{1}=-\frac{1}{4}\mathtt{ln}[1-\frac{2\lambda _{1}^{2}/(2\sqrt{\omega
^{2}-\lambda _{\beta }^{2}}+\lambda _{3}/N)+\lambda _{3}}{(\omega
_{1}-\lambda _{3}/2)}].
\end{equation}

\end{document}